\documentclass[%
mathleft,%
final,%
]{an}
\usepackage{graphicx}
\usepackage{times}
\newcommand{\kms}{km\,s$^{-1}$}
\newcommand{\ms}{m\,s$^{-1}$}

\newcommand{\hhh}{\mbox{H$_2$ }}
\newcommand{\hhhns}{\mbox{H$_2$}}
\newcommand{\ma}{\mbox{m\AA}}
\newcommand{\dmu}{\mbox{$\Delta\mu/\mu$ }}

\begin{document}

\Pagespan{1}{}
\Yearpublication{2014}%
\Yearsubmission{2013}%
\Month{1}%
\Volume{335}%
\Issue{1}%
\DOI{This.is/not.aDOI}%

\title{Constraints on variations of m$_\mathrm{p}$/m$_\mathrm{e}$ based on UVES observations of \hhh}

\author{M. Wendt\inst{1}\fnmsep\thanks{Corresponding author.
  \email{mwendt@astro.physik.uni-potsdam.de}}
}
\titlerunning{Constraints on variation of \dmu}
\authorrunning{M. Wendt}
\institute{Institute of Physics and Astronomy, University Potsdam, 14476 Potsdam, Germany}

\received{30 Aug 2013}
\accepted{Jan 2014}
\publonline{later}

\keywords{cosmology: observations, quasars: absorption lines, early universe.}

\abstract{This article summarizes the latest results on the 
proton-to-electron mass ratio $\mu$ derived from \hhh observations at high redshift 
in the light of possible variations of fundamental physical constants.
 The focus lies on UVES observations of the past years as enormous progress 
was achieved since the first positive results on \dmu were published. With the better
understanding of systematics, dedicated observation runs, and numerous approaches to improve 
wavelength calibration accuracy, all current findings are in reasonable good agreement with no 
variation and provide an upper limit of \dmu $<  1\times 10^{-5}$ for the redshift
range of $2 <$ z $< 3$.}

\maketitle

\section{Introduction}
The Standard Model of particle physics  contains    several 
fundamental constants whose values cannot be predicted by theory and   
need to  be measured through experiments (Fritzsch \cite{Fritzsch09}).
 These are the masses of the elementary particles and the  dimensionless 
coupling constants which are assumed time-invariant  although   
in  theoretical models    which seek  to unify
the four forces of nature they  vary naturally on cosmological
scales. 
 The fine-structure constant $\alpha \equiv e^2/(4\pi \epsilon_0 \hbar c)$ and 
the proton-to-electron mass ratio, $\mu  = m_{\mathrm{p}} / m_{\mathrm{e}}$  are two constants that can
 be probed in the laboratory as well as  in the  Universe by means of 
observations of  absorption lines due to intervening systems in the spectra
 of distant quasars (QSO\footnote{Historically from 'quasi stellar objects'.}) and have  been 
 subject of numerous studies. 
The former is related to the electromagnetic force while the latter  
is  sensitive primarily   to
the quantum chromodynamic scale (see, i.e., Flambaum \cite{Flambaum04}). 

In principle, the $\Lambda_{{\mathrm{QCD}}}$  scale
should vary considerably faster than that of quantum electrodynamics
$\Lambda_{{\mathrm{QED}}}$. Consequently, the change in the proton-to-electron mass ratio,
 if any, should be larger than that of the fine structure constant.   
Hence,\,\,$\mu$  is an ideal candidate to search for possible cosmological
variations of the fundamental constants.

A probe of the variation of $\mu$ could be obtained by comparing
relative frequencies of the electro-vibro-rotational lines of \hhh
as first applied by Varshalovich \& Levshakov (\cite{Varshalovich93})
 after Thompson (\cite{Thompson75})
 proposed the general approach to utilize molecule transitions for $\mu$-determination.
The original paper by Thompson (\cite{Thompson75}) did not take into account the different
sensitivities within the molecular bands, which is the key of the modern approach. 
\begin{figure}
\includegraphics[angle=270, width=\linewidth]{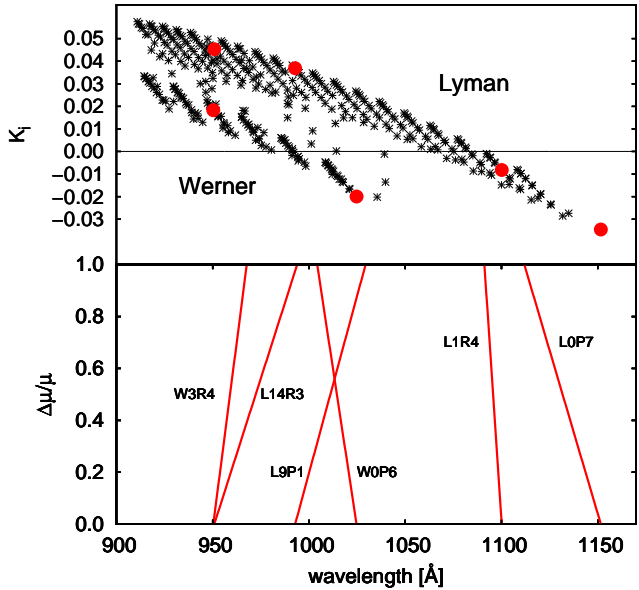}
\caption{The upper panel shows the sensitivity coefficients $K_i$ of the Lyman and Werner
transitions of \hhh against the restframe wavelength. Note, that the coefficients show 
different signs. The lower panel demonstrates the shifts of six selected transitions 
(marked with large red circles in the upper panel) with increasing \dmu. The transitions
referred to as L9P1 and W0P6
 even swap their positions in the observed spectra with 
increasing \dmu. For this illustration the range of \dmu is five orders of magnitude larger
than the current constraints on \dmu (see section \ref{sec:results}).}
\label{fig:ki}
\end{figure}
The method is based on the fact that the wavelengths of vibro-rotational
lines of molecules depend on the reduced mass, M, of the molecule.
For molecular hydrogen M $= m_{\mathrm{p}}/2$ so that the comparison of an observed vibro-rotational 
spectrum at high redshift with its present analog will  give information on the variation of
 $m_{\mathrm{N}}$ and $m_{\mathrm{e}}$.
Comparing electro-vibro-rotational lines with different
sensitivity coefficients gives a measurement of $\mu$.

The observed wavelength $\lambda_{\mathrm{obs},i}$  of any given
line in an absorption system at the redshift $z$ differs from the local
rest-frame
wavelength $\lambda_{0,i}$  of the same line in the laboratory according to the
relation 
\begin{equation}\label{eq:1}
\lambda_{\mathrm{obs},i}  =  \lambda_{0,i} (1 + z)\left(1+ K_i \frac{\Delta \mu}{\mu}\right),
\end{equation}   
where $K_i$ is the sensitivity coefficient of the $i$th component computed
theoretically for the Lyman and Werner bands of the \hhh
molecule (Meshkov et al. \cite{Meshkov07},
Ubachs et al. \cite{Ubachs07}).
Figure \ref{fig:ki} plots these sensitivity coefficient $K_i$ for the Lyman\footnote{The first digit
of an \hhh identifier indicates a Lyman or Werner line, followed by the vibrational quantum number
of the excited state and the branch with the rotational number, also referred to as $J$.} and Werner
transitions of \hhh in the upper panel. The coefficients are typically on the order
of $10^{-2}$, some coefficients differ in sign, which means that under the assumption
of positive variation of $\mu$, some \hhh lines are shifted into opposite directions.
The lower panel demonstrates this effect according to equation \ref{eq:1}.
The shifts for a strongly exaggerated \dmu of six lines are shown. The corresponding
sensitivity coefficients are marked as {\it red circles} in the upper panel.
The expected shifts at the current level of the constraint on \dmu are on the order of a 
few 100 \ms\, or about 1/10$^{\mathrm{th}}$ of a pixel size.

It is useful to measure  variations in velocities with comparison to the redshift
 of a given system defined by the redshift position of the lines with $K_i \approx0$,
 then introducing the reduced redshift  $\zeta_i$:
\begin{equation}
\zeta_i \equiv \frac{z_i - z}{1+z}
 = K_i \frac{\Delta\mu}{\mu}. \label{eq_LBLFM}
\end{equation}
 The velocity shifts of  the lines are   linearly proportional
 to $\Delta\mu/\mu$ which  can be measured  through a regression analysis in the
 $\zeta_i - K_i $ plane. 
This approach is referred to as line-by-line analysis in contrast to the 
comprehensive fitting method (CFM). The validity of these two approaches depends
mostly on the analyzed \hhh system. For example, the absorption in 
QSO 0347-383 (see Wendt \& Molaro \cite{Wendt12}) has the particular advantage of comprising 
a single velocity component, which renders observed transitions independent of each other
and allow for this regression method. This was also tested in
 Rahmani et al. (\cite{Rahmani13}) and King et al. (\cite{King08}).
For absorption systems with two or more closely and not properly resolved velocity components 
 many systematic errors may influence 
distinct wavelength areas. The CFM  fits all \hhh components along with additional H\,{\sc i}
 lines and introduces an artificially applied $\Delta\mu/\mu$ as free parameter in the fit. 
 The best matching  $\Delta\mu/\mu$ is then derived via the resulting $\chi^2$ curve.
 The CFM aims to achieve the lowest possible reduced $\chi^2$ via additional velocity components.
 In this approach, the information of individual transitions is lost because merely the
overall quality of the comprehensive model is judged. 
Weerdenburg et al. (\cite{Weerdenburg11}) increased the number of velocity components 
as long as the composite residuals of several selected absorption lines 
differ from flat noise. 
As pointed out by King et al. (\cite{King11}), for multi-component structures with overlapping velocity
components the errors  in the line centroids are heavily correlated and a simple $\chi^2$ regression
is no longer valid. The same principle applies for co-added spectra with relative velocity shifts.
The required re-binning of the contributing data sets introduces
 further auto-correlation of the individual 'pixels'. 
Rahmani et al. (\cite{Rahmani13}) discuss the assets and drawbacks of these two approaches in greater detail and
find evidence for an unresolved additional velocity component in the spectra for the QSO with the identifier
HE 0027 which renders the CFM favorable for that system.
The selection criteria for the number of fitted components are non-trivial and under debate.
Prause \& Reimers (\cite{Prause13}) discuss the possibility of centroid position shifts due to incorrect line decompositions
with regard to the variation of the finestructure constant $\alpha$, which in principle is applicable to
any high resolution absorption spectroscopy.
The uncertainties of the oscillator strengths $f_i$ that are stated to be up to 50\%
 (Weerdenburg et al. \cite{Weerdenburg11}) might further affect the choice
 for additional velocity components. 
Another source of deficient fits may be traced back to the nature of the bright background quasar
which in general is not a point-like source. In combination with the potentially
small size of the absorbing clumps of \hhhns, we may observe saturated absorption profiles with 
non negligible residual flux of quasar light not bypassing the \hhh cloud 
 (see Ivanchik et al. \cite{Ivanchik10}).

\section{Observations of extragalactic \hhh}
\begin{figure}
\includegraphics[angle=0, width=\linewidth]{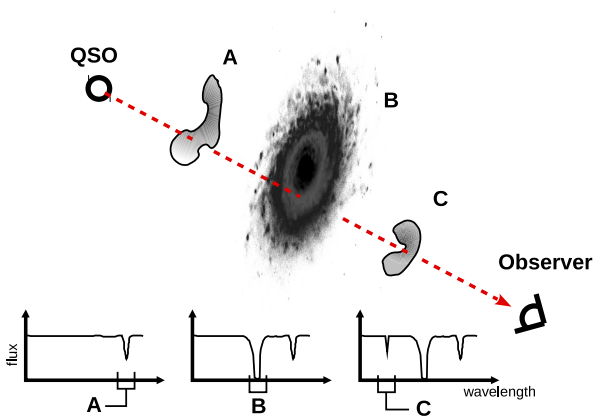}
\caption{Exemplary schematic for the line of sight of an observed
quasar. The absorbers {\tt A}, {\tt B} and {\tt C} are passed at decreasing redshift,
 respectively, and hence leave their signature at different redshifted
wavelengths in the observed spectrum. At {\tt B} the line of sight pierces
a dense part of a galaxy with more than $2\times10^{20}$
 atoms/cm$^2$ integrated, a so-called damped Ly-$\alpha$ (DLA) system. Only here 
extragalactic \hhh can be detected under special conditions.}
\label{fig:absorption}
\end{figure}
Molecular hydrogen is the most abundant molecule in the universe and plays
a fundamental role in many astrophysical contexts. It is found in all regions where the
shielding of the ultraviolet photons, responsible for the photo-dissociation
of \hhh, is sufficiently large.
Whenever \hhh is observed, it traces clumps of molecular hydrogen in galaxies, such as the MilkyWay 
(Savage et al. \cite{Savage77}) or regions nearby galaxies such as the Small Magellanic Cloud (SMC)
 or LMC (see, e.g.,
Richter et al. \cite{Richter98}; deBoer et al. \cite{deBoer98}). Observable amounts of \hhh can
 only be detected in the H\,{\sc i} disks of galaxies as pictured in Figure \ref{fig:absorption}
or in the the so-called Circum-Galactic Medium (CGM).

 Except in the very early universe, most \hhh is
likely to be produced via surface reactions on interstellar dust grains, since
gas-phase reactions are too slow in general (see, e.g., Harbart et al. \cite{Habart2004}).
The \hhh formation mechanism is
not yet fully understood. Direct observations of extragalactic \hhh are difficult since
electronic transitions occur only in the ultraviolet and mere roto-vibrational transitions
are forbidden due to the homo-nuclear nature of \hhh. This limits observations to
absorption systems at redshifts 2 to 3, for which the absorption features are shifted
into the optical. Another
problem is the narrow range of conditions under
which \hhh is formed. The required dust grains that allow for the forming of
\hhh can easily obscure the molecular hydrogen as well.

There are very few comprehensive studies on the spatial distribution of \hhh in
intervening absorption systems.
Hirashita et al. (\cite{Hirashita03}) computed the \hhh distribution based on 
 the ultraviolet background (UVB)
intensity and dust-to-gas ratios and find that \hhh has a very inhomogeneous,
 clumpy distribution on sub-parsec scale which explains the low number 
of known extragalactic \hhh absorbers.

Only a small fraction of observable lines of sight through DLAs cross
these isolated regions
(as illustrated in Figure \ref{fig:absorption}).
The Lyman bands in the wavelength range between 1000 \AA\, and 1100 \AA\, were first identified
in the absorption spectrum of a diffuse interstellar cloud
in the optical path towards $\xi$ Persei (Carruthers et al. \cite{Carruthers1970}).
Further
spaceborne observations also revealed absorption of Werner
bands and UV emission of Lyman and Werner
bands including their continua (Spitzer et al. \cite{Spitzer1974}).
Levshakov et al. (\cite{Levshakov1985}) tentatively assigned some
features in spectra obtained by Morton et al. (\cite{Morton1980}) from
PKS 0528-250 (one of the few systems up to date used for
determination of $\mu$, see table \ref{tab:h2systems}). Additional data of this system were taken
and used for a first constraint on a possible variation of $\mu$ put forward by Varshalovich \& Levshakov (\cite{Varshalovich93}).
\begin{table}
\caption[DLAs with \hhh absorption]{List of damped Lyman-$\alpha$ systems with
\hhh absorption observations with $z > 2$.}
\label{tab:h2}
\centering          
\begin{tabular}{l c c}  
\hline       
Quasar source & redshift $z_{\rm abs}$ & comment \\ 
\hline
J 2123-005     &	 2.06&\dmu measured\\
Q 1444+014   &        2.09&\\
Q 1232+082   &        2.34& \hhh saturated\\
Q 0841+129 & 2.37&\hhh very weak\\
Q 1439+113 & 2.42&\\
Q 2348-011  &	2.42&\dmu measured\\
HE 0027-184   &	2.42&\dmu measured\\
Q 2343+125   &        2.43&\hhh very weak\\
Q 0405-443   &        2.59&\dmu measured\\
FJ 0812+320 & 2.63& northern target\\
Q 0642-506 & 	2.66&\dmu measured\\
J 1237+064  &	2.69&\\
Q 0528-250   &        2.81&\dmu measured\\
Q 0347-383   &         3.02&\dmu measured\\
Q 1337+315 & 3.17& \hhh very weak\\
Q 1443+272   &         4.22& northern target\\
\hline
\label{tab:h2systems}
\end{tabular}
\end{table}
Observations with the Ultraviolet and Visual Echelle Spectrograph (UVES)
at the Very Large Telescope (VLT) in Paranal, Chile, led to \hhh detections towards
QSO 0347-383, QSO 1232+082 (Ivanchik et al. \cite{Ivanchik2002}; Levshakov et al. \cite{Levshakov02}),
 and towards
QSO 0551-336 (Ledoux et al.\cite{Ledoux2002}).
Additional observations of \hhh are reported
towards Q 0000-263 in Levshakov et al. (\cite{Levshakov2000}).
Ledoux et al. (\cite {Ledoux2003}) and Srianand et al. (\cite{Srianand2005}) performed
surveys on DLA systems at redshifts
$z > 1.8$, in which some new quasars with \hhh absorption
were detected.  
Later on, Petitjean et al. (\cite{Petitjean2006}) observed
the DLA systems in QSO 2343+125 and QSO 2348-011, while Ledoux et al. (\cite{Ledoux2006})
 observed \hhh lines in a spectrum at the highest redshift
to date with $z = 4.22$. In the course of the 'UVES Large Program for testing fundamental
physics' numerous QSOs were observed between 2010 and 2013 with the intention
to exploit the capabilities of the instrument and to achieve the best possible
constraints and variations of $\alpha$ and $\mu$ (see Molaro et al. \cite{Molaro13};
 Rahmani et al. \cite{Rahmani13}).
Noterdaeme et al. (\cite{Noterdaeme2008}) concluded from the comparison between
\hhhns-bearing systems and the overall UVES sample, that a significant increase
of the molecular fraction in DLAs could take place at redshifts $z_{\rm abs}
\geq 1.8$. The known
DLA systems with \hhh absorption in the redshift range observable with UVES  
and suitable for the determination of \dmu are listed in Table \ref{tab:h2}.

\section{Wavelength calibration for UVES}
In recent years the wavelength calibration emerged to be one of the most relevant aspects
in modern high resolution spectroscopy in particular for fields with such an accentuated
interest in line positions.
The error budget of the calibration is most difficult to estimate and its nature in some
aspects still unknown. 
Most high resolution UVES observations of QSOs for this purpose are taken with 
a slit width around 0.8\arcsec\, providing a resolving power in the order of 
$\lambda$/$\Delta \lambda$ $\approx 60\,000$\footnote{The resolving power R is 
not well defined and instrument specific (Robertson \cite{Robertson13}).}.
This value is to be understood as a rough estimate of the expected resolution as the
 resolving power  varies by up to 20 \% along individual wavelength orders
 (see Wendt \& Molaro \cite{Wendt12}).
The pixel-wavelength-conversion of the wavelength calibration  is done by using
 the corresponding  calibration spectrum. 
Murphy et al. (\cite{Murphy08}) and Thompson et al. (\cite{Thompson09a}) 
independently showed  that the
standard Th/Ar line list used in older UVES pipelines  was the primary
limiting factor. The laboratory wavelengths of the calibration spectrum were
only given to three decimal places (in units of \AA)
and, in many cases, the wavelengths were truncated rather than
rounded  from four decimal places (see Murphy et al. \cite{Murphy07}).

Thompson et al. (\cite{Thompson09b}) re-calibrated the wavelength solutions using
the calibration line spectra taken during the observations of the  QSOs and
argued that the new wavelength calibration was a key element in their null result.
They describe the whole calibration process for UVES in great detail.
In Wendt \& Molaro (\cite{Wendt12}), typical residuals of
the wavelength calibrations were $\sim$~0.34~\ma \,or $\sim$~24~\ms\,
at 4000 \AA. 
Malec et al. (\cite{Malec10}) give wavelength calibration residuals
of RMS $\sim$ 80 \ms\,
 for KECK/HIRES, and Rahmani et al. (\cite{Rahmani13}) obtained an
 average error of the wavelength solution of $\sim$ 40-50 \ms\,
which is the
 expected range for the UVES pipeline and ThAr lamp based
calibration in general.
However, this error estimate only accounts for the statistical aspect of the wavelength calibration.
Intrinsic systematics cannot be identified that easily.
Only calibration lamp exposures taken  immediately before and after 
object observations provide  an accurate monitoring  of  the physical  conditions.
 Moreover, the calibration frames were taken in special mode to avoid
 automatic spectrograph resetting at the
start of every exposure. Since Dec 2001 UVES
has implemented an automatic resetting of the Cross Disperser
encoder positions at the start of each exposure. This implementation has been
done to have the possibility to use daytime
ThAr calibration frames for saving night time. If this is excellent
for standard observations, it is a problem  for the measurement of fundamental
constants which requires  the best possible wavelength
calibration.
Additionally, thermal-pressure changes  move in the cross dispersers in different ways,
 thus introducing relative shifts between
the different spectral ranges in  different exposures.
There are no measurable temperature changes for the short exposures of
the calibration lamps but during the much longer science exposures the
temperature drifts generally by $\leq 0.2$ K, The estimates for
UVES are of 50 \ms\,  for $\Delta$T = 0.3 K or a $\Delta$P = 1 mbar
(Kaufer et al. 
\cite{Kaufer2004}), thus assuring a radial velocity
stability within $\sim 50$ \ms.
The motion of Earth during observation smears out the line by $\pm$ 40 \ms,
since the line shape itself remains symmetric, this does not directly impact
the centroid measurements but it will produce an absorption profile
that is no longer  strictly Gaussian (or Voight) but rather
slightly squared-shaped which further limits the quality of a line fit and must be
considered for multi-component fits of high resolution spectra.

The measured air wavelengths are commonly converted to vacuum via 
the dispersion formula by Edlen (\cite{Edlen66}). 
Drifts in the refractive index of air inside the spectrograph between the
ThAr and quasar exposures will therefore cause mis-calibrations.
According to the Edlen formula for the refractive index of air,  temperature and atmospheric
 pressure changes  of  1 K and 1 mbar  would cause differential velocity shifts in the optical
 in the order of $\sim 10$ \ms. 

A stronger concern is the possibility of much larger 
distortions within the spectral orders which  have been investigated at the
Keck/HIRES spectrograph by comparing the ThAr wavelength
scale with a second one established from I2-cell observations of a bright
quasar by Griest et al. (\cite{Griest10}). They find absolute offsets 
which can be as large as $500 - 1000$ \ms\,  and an 
additional distortion of about $300$ \ms\, within the individual orders.
The distortion is such that transitions at the edges of the Echelle orders
appear at different velocities with respect to transitions at the order
 centers when calibrated with a ThAr exposure. 
This would introduce relative velocity shifts between
different absorption features up to a magnitude the analysis
 with regard to $\Delta\mu/\mu$ is sensitive to.
 Whitmore et al. (\cite{Whitmore10}) repeated the same test for UVES with similar results
though the distortions fortunately show lower peak-to-peak velocity
variations of $\sim 200$ \ms, and Wendt \& Molaro (\cite{Wendt12}) detected indications for this
effect directly in the measured positions of \hhh features as well.

Molaro et al. (\cite{Molaro11}) suggested to use the available solar lines atlas in combination
with high resolution asteroid spectra taken close to the QSO observations to
 check UVES interorder distorsions and published
a revised solar atlas in Molaro \& Monai (\cite{Molaro12}).
Such asteroid spectra were used as absolute
calibration to determine velocity drifts in their data via cross-correlation of individual wavelength
intervals in Rahmani et al. (\cite{Rahmani13}).
They found distinct long range drifts of several $100$ \ms\, within 1000 \AA.
The origin of these drifts remains currently unknown but is verified 
in Whitemore et al. (2013 in prep.) and considered
by Bagdonaite et al. (\cite{Bagdonaite13b}) to contribute to their positive signal.
Rahmani et al. (\cite{Rahmani13}) found the drift to be constant over a certain epoch and
applied suitable corrections. So far the drift, when present, in UVES spectra always showed the
same trend at different magnitudes which could be an explanation for the reported tendency towards
positive variation in $\mu$ (see section \ref{sec:results}).

\section{Determination of precise line centroids}
\begin{figure}
\includegraphics[angle=270, width=\linewidth]{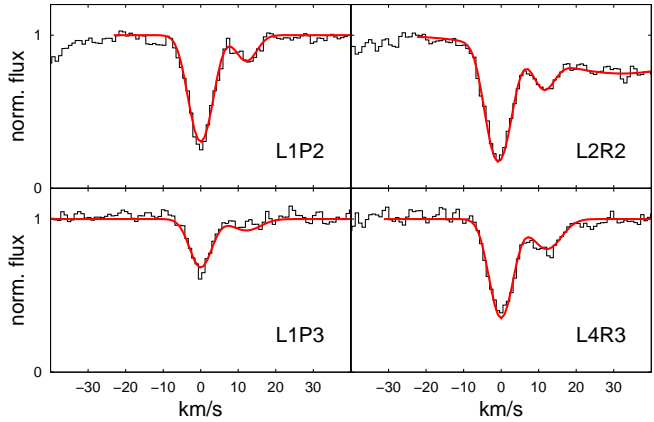}
\caption{Four exemplary \hhh features in QSO 0405-443 to demonstrate the tied fitting parameters.
Given in each panel at the lower right is the identifier of the observed transition. An additional
broad Ly-$\alpha$ component was added for L2R2, the \hhh is fitted with two velocity components with a
separation of 12.2 \kms. The shown fit is based merely on these four features.}
\label{fig:h2}
\end{figure}
Table \ref{tab:h2} lists all extragalactic \hhh features in DLA systems known today that are in principle
suitable for \dmu measurements.
Those already analyzed for the purpose of \dmu determination are pointed out in the rightmost column.
Q2348-011 shows a quite complex velocity structure in \hhh including detected self-blending which
is partially
reflected in the comparably large errorbars on this measure (see Bagdonaite et al. \cite{Bagdonaite12}).
An important benefit from analyzing molecular hydrogen in DLA systems is the large number of
observable transitions which lies in the order of 50-100 for the observed systems.
Considering the very limited typical size of the dense \hhh absorbers on the (sub)parsec scale,
we are confident that all observed transitions occur in the same physically bound medium.
This information can be used in simultaneous fits of all observed \hhh features with tied
parameters. In general a common column density is fit to all transitions from the same rotational level.
 At sufficiently high resolution it becomes mandatory to take into account the possible
multiphase nature of different J-levels as proposed in
 Noterdaeme et al. (\cite{Noterdaeme07}) and applied, i.e., 
in Wendt \& Molaro (\cite{Wendt12}) and Rahmani et al. (\cite{Rahmani13}), who fitted individual 
line broadening parameters to different $J$-levels.

QSO 0405-443 shows a weaker second velocity component which is well resolved.
Figure \ref {fig:h2} shows four transitions, namely L1P2, L2R2, L1P3, and 
L4R3.
 These features
were fitted with a common column density for each component in the $J=2$ transitions in the panels 
and one for the $J=3$ lines. To each $J$-group the b-parameters per velocity component were coupled
and a global velocity separation of both components was fitted to all four lines.
The small number of simultaneously fitted lines shows in the lower panel for the second component
whose column density for L1P3 was probably overestimated due to some contamination of L4R3.
Only for the typically large number of lines per rotational level these effects cancel out.
Former studies of QSO 0405 neglected the second component at an offset of about 12 \kms\,
 for the 
analysis of \dmu as it 
is considerably weaker. However, it can be utilized to get a proper measure of the oscillator
strengths of the transitions, which are stated to differ notably from the calculated values
 (see Weerdenburg et al. \cite{Weerdenburg11}). For this case the oscillator strengths 
were allowed to vary by up to 20 \% which resulted in improved fits. The additional, resolved
second component breaks  the degeneracy for limited signal-to-noise data of the
 column density $N$,
broadening paramter $b$, and oscillator frequency $f_{\mathrm{osc}}$ as the latter of course
is identical for all velocity components. These particular entangled fits are carried out with
an evolutionary fitting algorithm, that allows complex links between fitting parameter.
The routine is based on Quast, Baade \& Reimers (\cite{Quast05}) and was verified for this 
application in Wendt \& Reimers (\cite{Wendt08}).
An analysis making use of this additional boundary condition for simultaneous fits is currently
in progress (Wendt et al. 2014 in prep.).

\section{Published results for \dmu since 2008 based on UVES observations of \hhh}
\label{sec:results}
\begin{figure}
\includegraphics[angle=0, width=\linewidth]{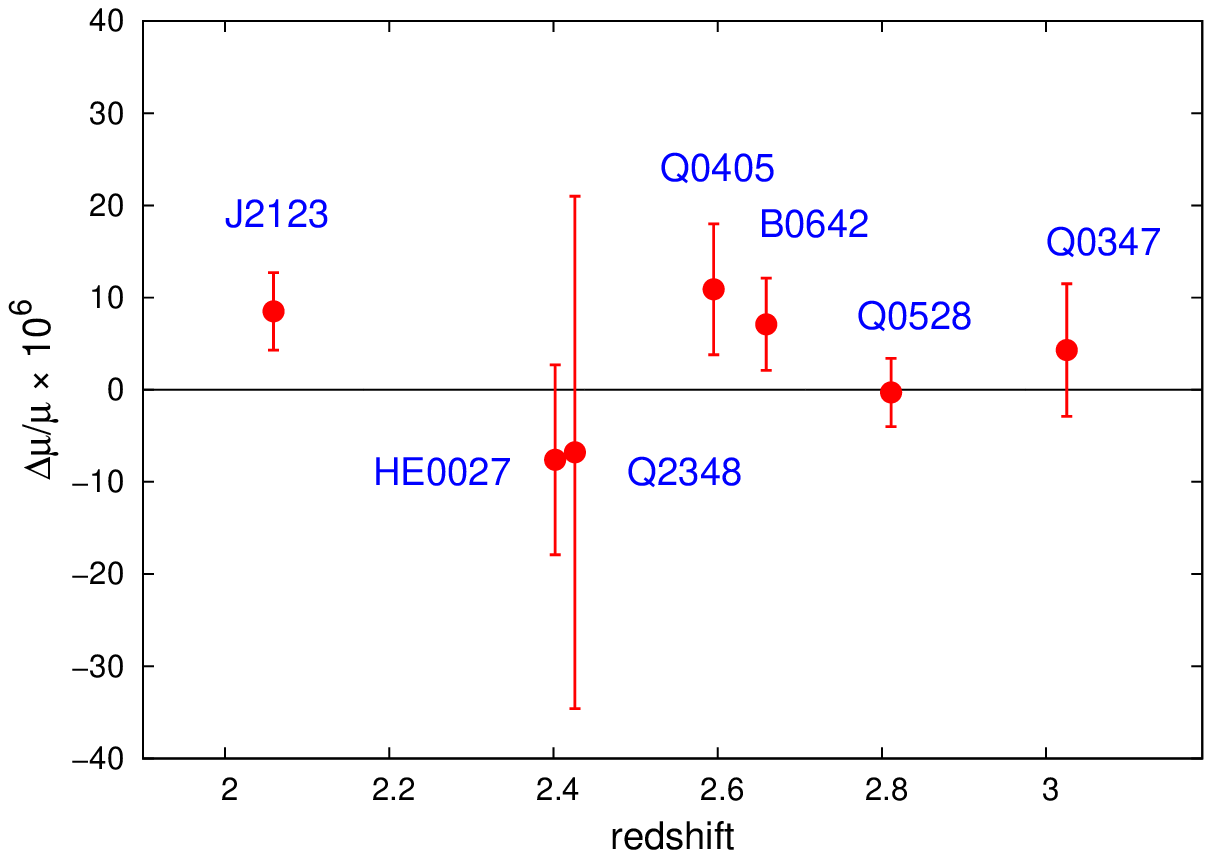}
\caption{Latest  results for \dmu based on \hhh in seven different quasar sightlines observed with UVES:
J2123-0050 (van Weerdenburg et al. \cite{Weerdenburg11}), HE0027-1836 (Rahmani et al. \cite{Rahmani13}),
Q2348-011 (Bagdonaite et al. \cite{Bagdonaite12}), Q0405-443 (King et al. \cite{King08}),
B0642-5038 (Bagdonaite et al.\cite{Bagdonaite13b}), Q0528-250 (King et al. \cite{King11}),
Q0347-383 (Wendt \& Molaro \cite{Wendt12}). The given errorbars are the sum of statistical and systematic error
(if both are given) under the assumption of gaussian distributed errors.
}
\label{fig:results}
\end{figure}
There were numerous determinations of \dmu  ever since early indications 
of change in $\Delta\alpha/\alpha$ were published and in particular after the paper by Reinhold et al.
 (\cite{Reinhold06}) who presented evidence that the proton-to-electron
mass ratio was larger in the past at the $> 3.5 \sigma$ level.
Several inconsistencies between different analyses for \dmu and $\Delta\alpha/\alpha$
 surfaced until it became evident that systematic
errors were responsible for the large scatter in the results.
The following Table \ref{tab:h2observations} lists publications 
that determined \dmu based on observations of \hhh at high redshift with UVES.
For clarity, only works since 2009 (with one exception) are listed.
This also rectifies the situation of cited earlier results that likely were dominated 
by the mentioned systematics in the wavelength calibration and limited accuracy of
 the restframe wavelengths as well as insufficient data quality.
King et al. (\cite{King08}) is added to the list since they determined the latest published 
value for \dmu for QSO 0405-443, which is currently being re-analyzed.
The value for \dmu based on J 2123-0050 in van Weerdenburg et al.	(\cite{Weerdenburg11})  was 
preceded by Malec et al.	(\cite{Malec10}) who observed the same system with the KECK telescope. 
Both results are in reasonable good agreement.
The publications in the table are sorted by the time the data were taken, which
is loosely correlated to the data quality with regard to the calibration preparations and
general high resolution and signal-to-noise requirements for the purpose of probing fundamental
physics.

Figure \ref{fig:results} shows the latest measurements of \dmu based on \hhh observations with UVES
for 7 observed quasar spectra. 
Bagdonaite et al. (\cite{Bagdonaite13b}) raise the point that they found evidence based on asteroid 
observations for a possible long range distortion of about 350 \ms\, over 1000 \AA\, in their observations
of B0642, a systematic that would
 translate to an offset in \dmu of -10 $\times 10^{-6}$. The analyzed asteroid spectra, however, were 
not taken during the observations of the quasar and hence were not applied as correction.
The presented data of seven measurements yields a mean of \dmu = ( 3.7 $\pm$ 3.5 ) $\times$ 10$^{-6}$ 
and is in good agreement with a non-varying proton-to-electron mass ratio \footnote {If the mentioned
 correction is applied,
this changes to \dmu = ( 2.3 $\pm$ 2.8 ) $\times$ 10$^{-6}$.}.
Such a generic mean value does not take into account any interpretation with regard to spatial or temporal
 variation and instead merely evaluates the competitive data available for \dmu based on \hhhns-observations, 
which consequently are limited to the redshift range of $2 <$ $z_\mathrm{abs} < 3$ and the evidence
these data provide for any non constant behavior of $\mu$ over redshift.
This tight constraint already falsifies a vast number of proposed theoretical models for varying $\mu$ or $\alpha$.
Thompson et al. (\cite{Thompson2013}) come to the conclusion that that ``adherence to the measured invariance in $\mu$ is a
 very significant test of the validity of any proposed cosmology and any new physics it requires''.

Current efforts to \dmu measurements exploit the technical limits of the UVES spectrograph and bring forward 
valuable new aspects of quasar absorption system analysis. 
The data from the ESO LP\footnote{ESO telescope program L185.A-0745} observations has the potential to set a new cornerstone in the assessment
of variability of fundamental physical constants. Observations featuring instruments in the foreseeable
future will provide further insights and allow to bring cosmological measurements of intergalactic
absorbers into a new era. Spectra recorded via laser-comb calibrated spectrographs (such as CODEX or EXPRESSO)
at large telescopes (E-ELT or VLT, respectively) implicate new methods of data analysis as well.
High resolution spectroscopy gains importance in several fields.
The next generation of instruments, 
such as the Potsdam Echelle Polarimetric and Spectroscopic Instrument (PEPSI) designed for the 
at the Large Binocular Telescope (LBT) features resolutions
 of up to 310\,000 (Strassmeier et al. \cite{Strassmeier2008}) and will already require improvements of
absorption line modeling from simple symmetric Voigt-profiles to more physical models incorporating
intrinsic velocity structures and inhomogeneities of the absorber.
\begin{appendix}
\renewcommand{\arraystretch}{1.5}
\begin{table*}
\caption{\dmu measurements based on \hhh observations with VLT/UVES.}\label{tab:h2observations}
\begin{tabular}{lccccc}
\hline
Reference & object & data & redshift & \dmu $\times 10^{-6}$ & comment\\
\hline
King et al. (\cite{King08}) & QSO 0405-443 & (2002) & 2.595 & 10.9 $\pm$ 7.1 & \parbox[t]{4cm}{data from earlier analyses\\new calibration applied} \\
Thompson et al. (\cite{Thompson09a}) & QSO 0347-383 &(2002) & 3.025 & -28 $\pm$ 16 & \parbox[t]{4cm}{thorough new calibration\\ ThAr linelist renewed}\\
Wendt \& Molaro (\cite{Wendt11}) & QSO 0347-383	& (2002,2003)& 3.025 &15.0 $\pm$ 10.8 & \parbox[t]{4cm}{utlizing independent data from 2003 to gain robustness}\\ 
Ubachs	et al. (\cite{Ubachs11}) & QSO 2348-011 & (2007) & 2.426 &-15 $\pm$ 16 & \parbox[t]{4cm}{non-ideal data\\ improved in 2013}\\
Malec	et al.	(\cite{Malec10}) & J 2123-0050	& (2007) & 2.059 & 5.6 $\pm$ 6.2& \parbox[t]{4cm}{based on KECK data}\\
van Weerdenburg et al.	(\cite{Weerdenburg11}) & J 2123-0050 & (2008) & 2.059 & 8.5  $\pm$  4.2 & \parbox[t]{4cm}{based on UVES data}\\
Bagdonaite et al.(\cite{Bagdonaite12})& QSO 2348-011 &	(2008)& 2.426 & -6.8 $\pm$ 27.8 & \parbox[t]{4cm}{complex \hhh system\\ refined analysis}\\
Bagdonaite et al.(\cite{Bagdonaite13b})& B 0642-5038 &	(2004,2008) & 2.659 & 17.1 $\pm$ 5.0 & \parbox[t]{4cm}{result possibly influenced by long range drift}\\
King et al. (\cite{King11}) &	QSO 0528-250 &	(2008,2009) &2.811 &-0.3 $\pm$ 3.7&\parbox[t]{4cm}{re-observation due to slit and calibration issues in former analysis, \hhh and HD detected}\\
Wendt \& Molaro (\cite{Wendt12}) & QSO 0347-383 & (2009) & 3.025 & 4.3  $\pm$  7.2&\parbox[t]{4cm}{re-observation to exploit UVES capabilities, LP specifications}\\
Rahmani et al. (\cite{Rahmani13}) &HE 0027-1836	& (2010-2012) & 2.402 &-7.6 $\pm$ 10.3&\parbox[t]{4cm}{LP data, quantitative determination of long range drift}\\
\hline
\end{tabular}
\end{table*}
\end{appendix}

\acknowledgements
  We appreciate the cooperation with the authors of AN and are thankful for the organizers
 of the 10$^{\mathrm{th}}$ Potsdam Thinkshop on 'High resolution optical spectroscopy' which motivated this article
on the status quo. We further want to express our gratitude to Philipp Richter and Helge Todt
for helpful comments and discussions.

\end{document}